\documentclass[conference,compsoc]{IEEEtran}
\usepackage{epsfig,endnotes}

\usepackage{booktabs} 
\usepackage{hyperref}

\newcommand{\COMMENTS}{no}

\usepackage[dvipsnames]{xcolor}
\definecolor{steel_blue}{RGB}{70, 130, 180}
\usepackage{amsmath}
\usepackage{pifont}
\usepackage[ruled, vlined]{algorithm2e}
\usepackage{fancybox}
\usepackage{pseudocode}
\usepackage{algpseudocode}
\usepackage{booktabs} 
\usepackage{outlines}
\usepackage{listings} 
\usepackage{tabularx}
\usepackage{colortbl, graphicx}
\usepackage{subcaption}
\usepackage{balance}
\usepackage{paralist}
\usepackage{tcolorbox}
\usepackage{wrapfig}

\usepackage{pifont}
\newcommand{\cmark}{\ding{51}}%
\newcommand{\xmark}{\ding{55}}%

\definecolor{darkgreen}{rgb}{0.0, 0.65, 0.0}

\ifthenelse{\equal{\COMMENTS}{no}}{%
\newcommand{\ag}[1]{\textit{\textcolor{blue}{[arpit]: #1}}} 
\newcommand{\rjh}[1]{\textit{\textcolor{darkgreen}{[rob]: #1}}} 

}
{
\newcommand{\ag}[1]{}
\newcommand{\lv}[1]{}
\newcommand{\rmnote}[1]{}
\newcommand{\rbnote}[1]{}
\newcommand{\nf}[1]{}
\newcommand{\mcnote}[1]{}
\newcommand{\rjh}[1]{}
\newcommand{\rb}[1]{}
}

\usepackage{xspace}
\newcommand*{\eg}{e.g.,\@\xspace}

\newcommand*{\ie}{\textit{i.e.,}\@\xspace}

\lstdefinelanguage{Python}
{keywords={MapBolt, ReduceBolt, >>, >+, +, &, push, if_, elif_, else, pop, 
match, fwd, modify, set, mod, sample, sampleD, sampleS, drop, $\triangleleft$, announce, 
withdraw, mapD, mapS, map, reduce, filter, filterD, filterS, runningReduceD, runningReduceS, 
distinct, toList, mapValues, countS, countByWindow, countByValueAndWindow, reduceByKey, window,
transform}, 
  sensitive=true, alsoletter={-,>>,+,&,|,_},comment=[l][\footnotesize\sffamily\textbf]{\!}
}

\lstdefinelanguage{p4}
{keywords=[2]{control, action, else, if, table, blackbox, register, field_list, field_List, field_list_calculation},
 otherkeywords={register_read, modify_field, apply, bit_or, register_write, input, algorithm, output\_width, width, default\_action, size, instance_count, exact,lpm,reads,actions  ,bit_and, clone_ingress_pkt_to_egress, add_header}, 
sensitive=true, alsoletter={-,>>,+,&,|,_},
basicstyle=\color{red}\ttfamily,
keywordstyle=\color{darkgreen}\ttfamily,
keywordstyle=[2]\ttfamily\bfseries\color{steel_blue},
commentstyle=\ttfamily, 
stringstyle=\ttfamily, 
identifierstyle=\ttfamily,
alsoletter={0,1,2,3,4,5,6,7,8,9}
comment=[l][\footnotesize\ttfamily\color{red}]{\!}
}

\lstdefinelanguage{query2}
{keywords={MapBolt, ReduceBolt, >>, >+, +, &, push, if_, elif_, else, pop, mapInit, 
match, fwd, modify, set, mod, sample, sampleD, sampleS, $\triangleleft$, announce, 
withdraw, mapD, mapS, map, reduce, filter, filterD, filterS, runningReduceD, runningReduceS, 
distinct, toList, mapValues, countS, countByWindow, countByValueAndWindow, reduceByKey, window,
transform, map-init, update-metadata, update-headers, emit, join}, 
keywordstyle=\bfseries\ttfamily,
keywordstyle=[2]\ttfamily\bfseries,
commentstyle=\ttfamily, 
stringstyle=\ttfamily, 
identifierstyle=\ttfamily,
emph={trafficAnomalyIPs}, 
emphstyle=\ttfamily\bfseries\color{red},
sensitive=true, alsoletter={0,1,2,3,4,5,6,7,8,9,-,>>,+,&,|,_},comment=[l][\footnotesize\sffamily\textbf]{\!}
}

\lstdefinelanguage{query3}
{keywords={MapBolt, ReduceBolt, >>, >+, +, &, push, if_, elif_, else, pop, mapInit, 
match, fwd, modify, set, mod, sample, sampleD, sampleS, $\triangleleft$, announce, 
withdraw, mapD, mapS, map, reduce, filter, filterD, filterS, runningReduceD, runningReduceS, 
distinct, toList, mapValues, countS, countByWindow, countByValueAndWindow, reduceByKey, window,
transform, map-init, update-metadata, update-headers, emit, join}, 
keywordstyle=\bfseries\ttfamily,
keywordstyle=[2]\ttfamily\bfseries,
commentstyle=\ttfamily, 
stringstyle=\ttfamily, 
identifierstyle=\ttfamily,
emph={map}, 
emphstyle=\ttfamily\bfseries\color{red},
sensitive=true, alsoletter={0,1,2,3,4,5,6,7,8,9,-,>>,+,&,|,_},comment=[l][\footnotesize\sffamily\textbf]{\!}
}

\lstdefinelanguage{Scala}%
  {morekeywords={abstract,case,catch,class,def,%
    do,else,extends,false,final,finally,%
    for,if,implicit,import,lazy,match,mixin,%
    new,null,object,override,package,%
    private,protected,requires,return,sealed,%
    super,this,trait,true,try,%
    type,val,var,while,with,yield},
otherkeywords={=,=>,<-,<\%,<:,>:,\#,@},%
   sensitive,%
   morecomment=[l]//,%
   morecomment=[n]{/*}{*/},%
   morestring=[b]",%
   morestring=[b]',%
   morestring=[b]""",%
  }[keywords,comments,strings]%

\usepackage{adjustbox}
\usepackage{lipsum}
\usepackage{amsthm}

\usepackage{collcell}
\usepackage{hhline}
\usepackage{pgf}
\usepackage{multirow}

\newcolumntype{R}[2]{%
    >{\adjustbox{angle=#1,lap=\width-(#2)}\bgroup}%
    l%
    <{\egroup}%
}

\def\colorModel{RGB} 

\newcommand\ColCell[1]{
  \pgfmathparse{#1}
  \pgfmathtruncatemacro\resultr{23+#1}
  \pgfmathtruncatemacro\resultg{55+#1}
  \pgfmathtruncatemacro\resultb{94+#1}

   \ifnum\pgfmathresult=0
        \relax\color{white}
        \pgfmathsetmacro\compA{255}      
        \pgfmathsetmacro\compB{255} 
        \pgfmathsetmacro\compC{255} 
    \else
        \definecolor{foo}{RGB}{\resultr,\resultg,\resultb}  
        \relax\color{foo}
        \pgfmathsetmacro\compA{\resultr}      
        \pgfmathsetmacro\compB{\resultg} 
        \pgfmathsetmacro\compC{\resultb} 
    \fi
  \edef\x{\noexpand\centering\noexpand\cellcolor[\colorModel]{\compA,\compB,\compC}}\x #1
  } 
\newcolumntype{E}{>{\collectcell\ColCell}m{0.4cm}<{\endcollectcell}}  

\newcommand{\stratonerescued}{\$3.4M\xspace}

\newcommand{\totalrescued}{\$11.2M\xspace}

\newcommand{\stratonebt}{\$114M\xspace}
\newcommand{\strattwobt}{\$296M\xspace}
\newcommand{\totalrescuedbt}{\$410M\xspace}

\newcommand{\stratonerescuedamt}{8\xspace}

\newcommand{\totalrescuedamt}{28\xspace}

\usepackage{listings}
\usepackage{xspace}

\graphicspath{{figure/}{figures/}}

\usepackage{amsmath}
\usepackage{pifont}

\newcommand{\sys}{\textsc{BackRunner}\xspace}

\newcommand{\tx}{\texttt{tx}}

\newcommand{\PP}[1]{
  \vspace{2px}
  \noindent{\bf \IfEndWith{#1}{.}{#1}{#1.}}
}

\usepackage{listings, xcolor}

\definecolor{verylightgray}{rgb}{.97,.97,.97}

\lstdefinelanguage{Solidity}{
	keywords=[1]{anonymous, assembly, assert, balance, break, call, callcode, case, catch, class, constant, continue, constructor, contract, debugger, default, delegatecall, delete, do, else, emit, event, experimental, export, external, false, finally, for, function, gas, if, implements, import, in, indexed, instanceof, interface, internal, is, length, library, log0, log1, log2, log3, log4, memory, modifier, new, payable, pragma, private, protected, public, pure, push, require, return, returns, revert, selfdestruct, send, solidity, storage, struct, suicide, super, switch, then, this, throw, transfer, true, try, typeof, using, value, view, while, with, addmod, ecrecover, keccak256, mulmod, ripemd160, sha256, sha3}, 
	keywordstyle=[1]\color{blue}\bfseries,
	keywords=[2]{address, bool, byte, bytes, bytes1, bytes2, bytes3, bytes4, bytes5, bytes6, bytes7, bytes8, bytes9, bytes10, bytes11, bytes12, bytes13, bytes14, bytes15, bytes16, bytes17, bytes18, bytes19, bytes20, bytes21, bytes22, bytes23, bytes24, bytes25, bytes26, bytes27, bytes28, bytes29, bytes30, bytes31, bytes32, enum, int, int8, int16, int24, int32, int40, int48, int56, int64, int72, int80, int88, int96, int104, int112, int120, int128, int136, int144, int152, int160, int168, int176, int184, int192, int200, int208, int216, int224, int232, int240, int248, int256, mapping, string, uint, uint8, uint16, uint24, uint32, uint40, uint48, uint56, uint64, uint72, uint80, uint88, uint96, uint104, uint112, uint120, uint128, uint136, uint144, uint152, uint160, uint168, uint176, uint184, uint192, uint200, uint208, uint216, uint224, uint232, uint240, uint248, uint256, var, void, ether, finney, szabo, wei, days, hours, minutes, seconds, weeks, years},	
	keywordstyle=[2]\color{teal}\bfseries,
	keywords=[3]{block, blockhash, coinbase, difficulty, gaslimit, number, timestamp, msg, data, gas, sender, sig, value, now, tx, gasprice, origin},	
	keywordstyle=[3]\color{violet}\bfseries,
	identifierstyle=\color{black},
	sensitive=true,
	comment=[l]{//},
	morecomment=[s]{/*}{*/},
	commentstyle=\color{gray}\ttfamily,
	stringstyle=\color{red}\ttfamily,
	morestring=[b]',
	morestring=[b]"
}

\usepackage{xcolor}

\definecolor{kscolor}{rgb}{0.9,0.1,0.1}
\definecolor{mscolor}{rgb}{0.1,0.1,0.9}
\definecolor{stcolor}{rgb}{0.1,0.9,0.1}

\definecolor{codegreen}{rgb}{0,0.6,0}
\definecolor{codegray}{rgb}{0.5,0.5,0.5}
\definecolor{codepurple}{rgb}{0.58,0,0.82}
\definecolor{backcolour}{rgb}{0.95,0.95,0.92}

\lstdefinestyle{mystyle}{
    backgroundcolor=\color{backcolour},   
    commentstyle=\color{codegreen},
    keywordstyle=\color{magenta},
    numberstyle=\tiny\color{codegray},
    stringstyle=\color{codepurple},
    basicstyle=\ttfamily\footnotesize,
    breakatwhitespace=false,         
    breaklines=true,                 
    captionpos=b,                    
    keepspaces=true,                 
    numbers=left,                    
    numbersep=5pt,                  
    showspaces=false,                
    showstringspaces=false,
    showtabs=false,                  
    tabsize=2
}

\definecolor{clcolor}{rgb}{0.5,0.7,0.9}
\definecolor{kscolor}{rgb}{0.9,0.1,0.1}
\definecolor{rbcolor}{rgb}{0.7,0.4,0.7}
\definecolor{nkcolor}{rgb}{0.4,0.7,0.7}
\definecolor{rfcolor}{rgb}{0.56, 0.0, 1.0}
\definecolor{cscolor}{rgb}{0.1, 0.4, 1.0}
\definecolor{jpcolor}{rgb}{0.36, 0.54, 0.66}

\begin{document}
\title{\sys: Mitigating Smart Contract Attacks in the Real World}

\author{
  \IEEEauthorblockN{
    Chaofan Shou\IEEEauthorrefmark{1},
    Yuanyu Ke\IEEEauthorrefmark{2},
    Yupeng Yang\IEEEauthorrefmark{3},
    Qi Su\IEEEauthorrefmark{2},
    Or Dadosh\IEEEauthorrefmark{4},
    Assaf Eli\IEEEauthorrefmark{4},
    David Benchimol\IEEEauthorrefmark{4},\\
    Doudou Lu\IEEEauthorrefmark{2},
    Daniel Tong\IEEEauthorrefmark{5},
    Dex Chen\IEEEauthorrefmark{5},
    Zoey Tan\IEEEauthorrefmark{5},
    Jacob Chia\IEEEauthorrefmark{2},
    Koushik Sen\IEEEauthorrefmark{1},
    Wenke Lee\IEEEauthorrefmark{3}
  }
  \IEEEauthorblockA{
    \IEEEauthorrefmark{1}University of California, Berkeley
  }
  
  \IEEEauthorblockA{
    \IEEEauthorrefmark{3}Georgia Institute of Technology
  }
    \IEEEauthorblockA{
    \IEEEauthorrefmark{2}Fuzzland Inc.
  }
  \IEEEauthorblockA{
    \IEEEauthorrefmark{5}Semantic Layer Labs
  }

  \IEEEauthorblockA{
    \IEEEauthorrefmark{4}Ironblocks
  }
  
}

\newcommand{\strategyone}{preemptive hijack\xspace}
\newcommand{\strategyonecapitalized}{Preemptive Hijack\xspace}
\newcommand{\strategytwo}{attack backrunning\xspace}
\newcommand{\strategytwocapitalized}{Attack Backrunning\xspace}











\sloppy
\pagestyle{plain}
\maketitle

\begin{abstract}
Billions of dollars have been lost due to vulnerabilities in smart contracts.
To counteract this, researchers have proposed attack frontrunning protections designed to preempt malicious transactions by inserting ``whitehat" transactions ahead of them to protect the assets.
In this paper, we demonstrate that existing frontrunning protections have become ineffective in real-world scenarios.
Specifically, we collected 158 recent real-world attack transactions and discovered that 141 of them can bypass state-of-the-art frontrunning protections.
We systematically analyze these attacks and show how inherent limitations of existing frontrunning techniques hinder them from protecting valuable assets in the real world.
We then propose a new approach involving 1) \textit{\strategyone}, and 2) \textit{\strategytwo}, which circumvent the existing limitations and can help protect assets before and after an attack.
Our approach adapts the exploit used in the attack to the same or similar contracts before and after the attack to safeguard the assets.
We conceptualize adapting exploits as a program repair problem and apply established techniques to implement our approach into a full-fledged framework, \sys.
Running on previous attacks in 2023, \sys can successfully rescue more than \totalrescuedbt.
In the real world, it has helped rescue over \totalrescued worth of assets in \totalrescuedamt separate incidents within two months. 
\end{abstract}

\section{Introduction}

Smart contracts on blockchain platforms have seen rapid growth and adoption over recent years. The immutability and transparency of blockchains make them well-suited for self-enforcing and self-executing programs, called smart contracts. However, these same properties also make vulnerabilities within smart contracts impactful, as malicious transactions cannot easily be reversed once executed. Numerous high-profile incidents of smart contract vulnerabilities being exploited for profit have resulted in billions of dollars worth of digital assets stolen or otherwise put at risk \cite{attack1,attack2,attack3,daohack,blocksec}.

Several defense mechanisms based on transaction frontrunning~\cite{yeim,xue2022preventing,qin2023blockchain} have emerged to curb the exploitation of smart contract vulnerabilities\footnote{Attack frontrunning is commonly confused with frontrunning attacks. Attack frontrunning refers to detecting and preventing attacks via frontrunning. The latter refers to attacks exploiting frontrunning vulnerabilities in the smart contract. This paper focuses on attack frontrunning.}. These techniques take advantage of the transparent nature of blockchains such as Ethereum to monitor transactions in the mempool that are waiting to be mined. By analyzing these pending transactions and the code of smart contracts they interact with, protective transactions can be constructed and prioritized to preempt malicious transactions. 

Although frontrunning techniques seem promising in theory, our research using honeypots and analyzing real-world measurement data reveals that they are still largely ineffective due to inherent limitations.
In the real world, almost all hackers leverage private transactions to hide from the public before the block is mined and broadcasted, making the attacks undetectable until they become hard to mitigate.
Moreover, implementing attack frontrunning techniques requires trivial effort, resulting in a high volume of malicious bots running simultaneously.
Consequently, when attacks appear in the public mempool, we discover that the bidding process among malicious bots and whitehat bots can sometimes lead to more than 80\% of the funds lost for rewarding the block builders.
Therefore, there is an urgent need for new alternative techniques capable of safeguarding funds from potential attacks effectively.

We propose a new strategy allowing us to rescue the funds before the attack.
We observe that most attacks are split into two steps: deploying the exploit contract and triggering the exploit. 
Existing techniques attempt to frontrun the second step without caring about the first one.
However, we found that the exploit in the first step provides valuable information about the attack.
Even though attackers may not put all the details of the attacks in the exploit (\eg configurations), we discover that traditional search-based program repair methodologies can effectively figure out the necessary missing information by treating them as "holes" and filling them.
Our strategy is to clone and mutate the exploit to make the whole attack successful and profitable to addresses we can control. 
Afterward, we can effectively counteract the attacks and "whitehat" hack the victims in minutes or even days before the actual attacks happen. 
We refer to this strategy as \textit{\strategyone}.

We also propose another strategy that mitigates potential losses after the attack.
We observe that attacks commonly leave residual risks, and the exploiter may not initially steal all assets from the victims. 
In the case of the INS20 hack\cite{ins20}, the initial attack only took \$2K, while the subsequent attacks led to more than \$692K loss. 
Additionally, most smart contract projects deployed have multiple deployments and forks on different chains. In the case of Curve Finance\cite{curve}, the initial attack targeting a smart contract led to \$11M loss, yet the subsequent attacks targeting other similar contracts led to more than \$60M loss. We observe that most exploits can be cloned and mutated (i.e., repaired) to target new victims. By automatically ``whitehat" hacking other potential victims, we can significantly reduce the potential loss after attacks. We refer to this strategy as \textit{\strategytwo}.

\begin{figure}
    \centering
    \includegraphics[width=8cm]{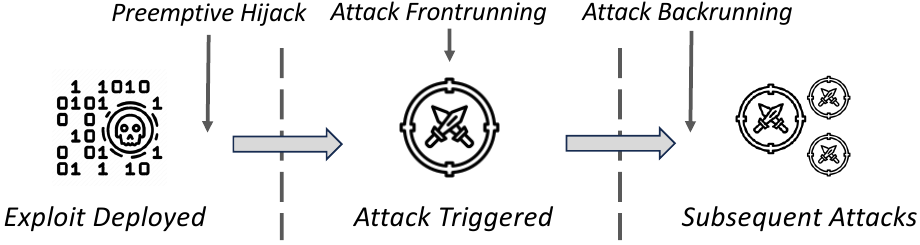}
    \caption{Time span of an attack}
    \label{fig:tar}
    \vspace{-5px}
    
\end{figure}

We depict \strategyone, \strategytwo, and their relations with attack frontrunning in \autoref{fig:tar}. 
Attack backrunning, compared to \strategyone, happens after the attack and has full information about the attack steps. 

We implement our strategies into a defense framework named \sys. Early testing of \sys has shown incredible promise for attack mitigation. Over two months of initial deployment, we successfully mitigated \totalrescuedamt real-world attacks and recovered over \totalrescued worth of assets that attackers would have taken otherwise.
In backtesting, our techniques were able to rescue more than \totalrescuedbt in assets in 2023. 

In summary, we make the following contributions:

\begin{enumerate}
    \item We demonstrate that attack frontrunning is ineffective based on real-world data analysis and experiments. 
    \item We propose two new defense strategies, \strategyone and \strategytwo, that can help mitigate real-world attacks and rescue funds. We also propose novel techniques to synthesize exploits for \strategyone and \strategytwo instantly and accurately.  
    \item We have implemented our strategies into a framework, \sys. Through backtesting, we demonstrate that \strategyone and \strategytwo work better than attack frontrunning in the real world. We conducted real-world deployment of \sys and rescued \totalrescued in a single month.  
\end{enumerate}



\section{Background}

\subsection{Smart Contracts}
Smart contracts\footnote{In the following sections, we refer smart contracts to be Ethereum Virtual Machine smart contracts, which can run on blockchains such as Ethereum, Binance Smart Chain, Polygon, etc. } are programs written in languages such as Solidity and Vyper that compile down to bytecode that runs on the Ethereum Virtual Machine (EVM). They are persistent scripts stored on the blockchain that allow developers to encode complex, self-executing logic. When a user submits a transaction that interacts with a contract, it triggers execution by the EVM, altering the contract's persistent state stored in the blockchain. Each computational step costs "gas" paid in small amounts of ether. Code execution only progresses as long as the gas limit set by the sender allows.

Smart contracts can access user inputs in the transactions and data such as the \verb|msg.sender| attributes to implement customized logic around blockchain transactions. For example, a contract can restrict functions only to authorize particular user addresses or require transactions to meet minimum ether value limits. Under the hood, contract storage works by mapping human-readable text variable names defined by the developer to 256-bit addresses in the permanent storage trie structure managed by the EVM. All computations and state alterations by a smart contract occur on this persisted data.

Attacks in smart contracts target vulnerabilities such as reentrancy issues, integer overflow or underflow errors, unprotected functions, reliance on external contracts, and more. One of the most infamous examples is the DAO hack on the Ethereum platform, where a reentrancy vulnerability allowed an attacker to repeatedly withdraw funds. 

Below, we introduce common services in blockchains such as Ethereum. These services are widely used and also leveraged by malicious exploits from the attackers. 
Familiarity with them can help understand the rest of the paper. 


\textbf{Tokens} Tokens are a key feature of Ethereum that enables the creation of digital assets and units of value on top of the Ethereum blockchain. These digital tokens are defined and managed through smart contracts. They can represent anything from virtual currencies and digital assets to voting rights or application access permissions. The ERC-20 standard~\cite{erc20} provides a common set of rules for defining fungible tokens on Ethereum. It specifies methods such as \verb|balanceOf()| to query an account's token balance, \verb|transfer()| to transfer tokens between accounts, and other functions to ensure consistent token behavior across different contracts. By conforming to the ERC-20 interface, tokens built on Ethereum can integrate seamlessly with exchanges, wallets, and other blockchain infrastructure designed around this standard. The standardized token interface facilitates issuing and distributing interoperable tokens with the broader Ethereum ecosystem. 

\textbf{Liquidity Pools (LP)} facilitate decentralized token trading on Ethereum through automated market maker (AMM) smart contracts\cite{uniswapv2}. A common implementation is Uniswap V2, where an LP holds reserves of two tokens and uses an algorithmic pricing formula to enable swaps between them. The AMM contract automatically sets prices according to the ratio of the quantities of the two tokens in the pool. This ratio determines the exchange rate between the pair based on the formula: $x * y = k$. Here, x and y are the token quantities and k is a constant. As trades occur, the balances change but the product stays equal to k, keeping the system in equilibrium. To trade tokens, users interact with the pool contract directly with no intermediaries. As trades shift the ratios, the pricing algorithm ensures prices adjust accordingly to maintain the constant k value. Liquidity providers supply reserve tokens to the pools to enable trading. In return, they earn fees from the trades occurring against those reserves. By automating swaps through programmatic supply-demand mechanisms, Uniswap and other AMMs allow fast, decentralized exchanges without order books or counterparties. 

\textbf{Flashloan}\cite{flashloan} service allows users to borrow substantial assets without any upfront collateral under the strict condition that the borrowed amount is returned within the same transaction. If the loan is not returned within it, the transaction is reverted as if it never occurred, ensuring the lender's assets are not at risk. This mechanism leverages the atomicity of blockchain transactions and has been utilized for various purposes, including arbitrage, collateral swapping, and debt refinancing. However, they have also been implicated in sophisticated DeFi attack vectors, as malicious actors can leverage flashloan to cause pricing discrepancies and conduct price manipulation attacks within a single transaction.

\subsection{Blockchain Mempool}
The mempool is the temporary holding area for transactions on the blockchain before they are included in a block\cite{memon2019simulation}. It operates as a queue, prioritizing transactions by the gas price. The concept of gas is critical to the mempool\cite{liu2022empirical}. Gas refers to the fee paid for executing transactions on the blockchain. Senders set a gas price they are willing to pay, which signals to miners the priority of that transaction. When the network gets congested, transaction senders increase their gas price to incentivize faster processing. This free market mechanism around gas pricing helps balance network capacity and usage. Senders set the priorities, and miners process based on profitability. This coordination through gas pricing allows the blockchain network to handle spikes in traffic and use. 

\subsection{MEV and Frontrunning}
The public visibility of pending transactions enables exploitation by Miner Extractable Value (MEV) bots, notably through frontrunning and backrunning \cite{torres2021frontrunner,daian2020flash,qin2022quantifying}. Since transactions in the mempool can be ordered based on gas fees paid, bots can monitor transactions and insert additional ones before and after target transactions to gain profits from arbitrage, liquidation, etc.

To facilitate these MEV bots and increase validator profits, protocols such as Flashbots~\cite{flashbot1,flashbot2,flashbot3} have separated the role of block building from that of validators. Specifically, dedicated block builders now organize and sequence transactions, optimizing orders for maximum fees or MEV profits. The block builders then transmit these optimized block layouts to validator nodes who propose the blocks.

Different block builders utilize various different policies~\cite{flashbot4}.
Some allow bundled transactions (i.e., an atomic sequence of transactions that no transaction is inserted in between) from MEV bots to remain intact for higher fees, while others receive direct payment from arbitrageurs in non-native tokens as an incentive. In all cases, block builders aim to maximize their own revenue share, creating the most profitable block organizations. The profit is ultimately divided between the proposing validator and the specialized block-builder. To maximize profits and compete with other block builders, when each block builder receives a transaction, the block builder does not propagate the transaction among the network but hoards it until the block is proposed. These transactions sent to block builders are known as private transactions. Currently, over 90\% of the blocks are being built by third-party block builders~\cite{flashbot5}, not by validators themselves.

\subsection{Smart Contract Firewall}

The concept of proactive defense against attacks was first proposed by a well-known security researcher OfficerCIA in 2021. Soon after that, BlockSec developed an attack frontrunning bot, cloning attack transactions and "whitehat" hacking victims, which successfully rescued more than \$15M assets since 2022 \cite{blocksec}. In the meantime, malicious actors have also recognized that frontrunning attacks is profitable. The first well-known occurrence of malicious attack frontrunning happened in Dec. 2022, targeting the AES project\cite{aes}. In 2023, Zhang et al.\cite{yeim} formalized the attack frontrunning technique. 

\subsection{Program Repair}
The concept of automatically repairing programs has existed for decades and has gained significant research interest in the last 10-15 years. Early work in this area focused on simple heuristic-based bug fixes or fixes tailored to domain-specific rules. Recently, techniques leveraging large language models, machine learning, formal methods, and program synthesis have produced more robust and general program repair solutions. Key techniques for automated repair include generate-and-validate\cite{gv0,gv1}, semantics-based analysis\cite{sem1,sem2,sem3}, program synthesis\cite{syn1}, machine learning-based repair\cite{repair1,repair2,repair3} models, and search-based software engineering\cite{search0,search1,search2}.



\section{Motivation}

\subsection{Attack Frontrunning}\label{motivation-atk-frontrunning}
\label{sec:motivation1}

We developed a smart contract firewall based on the state-of-the-art frontrunning techniques proposed in STING\cite{yeim} on the Ethereum and Binance Smart Chain networks for over six months. We encountered several fundamental limitations of frontrunning in this deployment, which demonstrated that frontrunning is ineffective in the current landscape. 
\begin{figure}
    \centering
    \includegraphics[width=8cm]{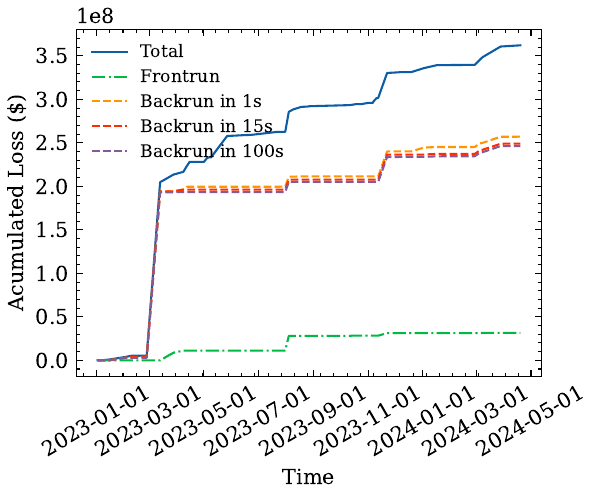}
    \caption{Funds loss in 2023 - 2024.  } 
    \label{fig:loss1}
\end{figure}
\begin{figure}
    \centering
    \includegraphics[width=8cm]{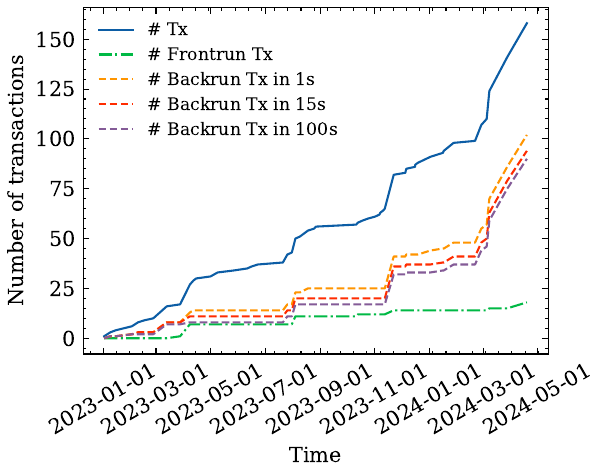}
    \caption{Successful attack transactions in 2023 - 2024.} 
    \label{fig:loss2}
    \vspace{-5px}
    
\end{figure}
To comprehensively assess the efficacy of frontrunning in mitigating attack impacts, we examined 158 documented attack transactions~\cite{blocksec} from 54 attacks that happened between 2023/01 and 2024/05, resulting in financial losses ranging from \$100K to \$200M. As shown in \autoref{fig:loss1} and \autoref{fig:loss2}, out of these attack transactions, only a small fraction (18 transactions) were intercepted by attack frontrunning bots, which rescued less than \$31.5M (8.7\% of the total loss) in assets, indicating a limited success rate in asset recovery through frontrunning.

Furthermore, to deepen our understanding of the frontrunning process, we collected and analyzed the mempool data, including the initial detection time of transactions by our nodes and those managed by Blocknative\cite{blocknative} globally. This investigation showed that merely 17 out of the 158 documented attack transactions were visible to the public before block broadcast, thereby allowing bots to engage in frontrunning. Most of the remaining attacks were executed using strategies that circumvented public visibility, such as employing block builders for sending private transactions.

\begin{figure}
    \centering
    \includegraphics[width=8cm]{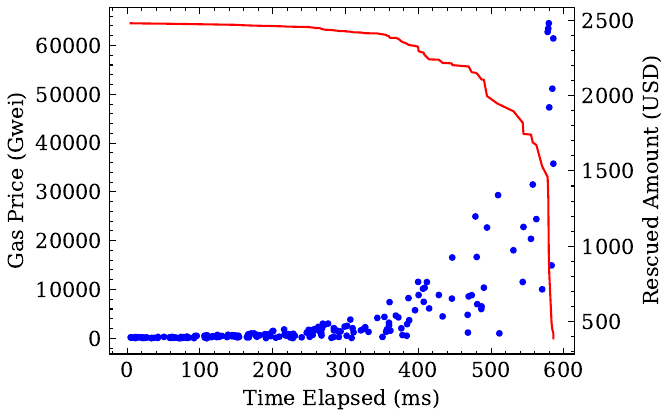}
    \caption{Block bidding process of frontrunning the \$3000 honeypot attack.}
    \label{fig:bid}
    \vspace{-5px}
\end{figure}

\vspace{5px}
\noindent\fbox{\begin{minipage}{23.5em}
\textbf{Observation 1}: Prevalence of private transactions have greatly reduced effectiveness of attack frontrunning.
\end{minipage}}

\vspace{5px}

Additionally, we discover that among these 17 attacks, more than 30\% of rescued funds are sent to block builders or paid as gas fees. The bots do so to ensure their transactions are placed before other frontrunning bots. To visualize the competition of bots on the network, we deployed a honeypot contract on Binance Smart Chain. In September 2023, we intentionally launched a public attack transaction to steal \$3K worth of assets in the honeypot contract and monitor the mempool. The bidding over time is visualized in \autoref{fig:bid}.
We observed that at least 6 bots had generated relevant transactions in an attempt to frontrun our attack transaction. These 6 bots competed with each other by continuously bidding higher gas prices in less than 600 milliseconds. In total, we observed 189 bids, with gas prices growing from 10 gwei to 60,000 gwei. As the gas price reaches 60,000 gwei, 80\% of the rescued funds are burnt and used to pay the validators, resulting in less than \$500 worth of assets rescued.

\vspace{5px}
\noindent\fbox{\begin{minipage}{23.5em}
\textbf{Observation 2}: Competitions between attack frontrunning bots lead to high gas prices, greatly reducing the funds that can be rescued. 
\end{minipage}}
\vspace{5px}

\begin{figure}
    \centering
    \includegraphics[width=8cm]{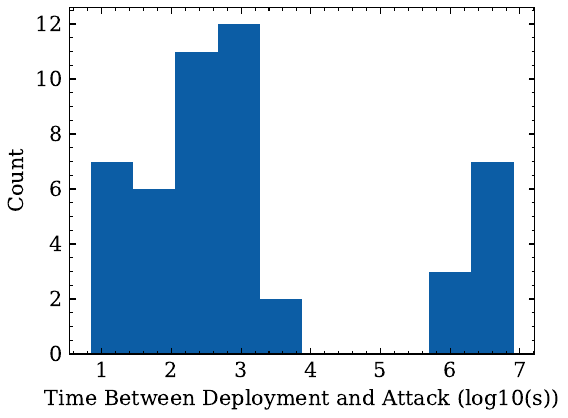}
    \caption{The time difference between when the exploit is deployed and the attack is triggered.}
    \label{fig:timediff}
    \vspace{-5px}
\end{figure}

Due to Observations 1 and 2, it is practically hard to frontrun an attack today as it is either impossible due to invisible private transactions or comes with a great cost due to the competition. 
In fact, we realize that attack frontrunning shall happen before the attack transaction is sent.
At first glimpse, this seems equivalent to performing vulnerability hunting on the chain, which is ineffective and hard to automate. 
However, we notice that before an attack, there are often many indicators that can provide us with attack information, such as exploit contract deployment or attacks on similar contracts. 
These indicators may enable us to identify potential victims and thus synthesize a counter-exploit.
Based on this insight, we design our first strategy, \textit{\strategyone}, which synthesizes counter-exploit from deployed exploits, and the second strategy, \textit{\strategytwo}, which adapts attack transactions to target similar contracts before potential residual attacks.

We demonstrate the effectiveness of both strategies by first conducting a statistical analysis of attacks in 2023.
We assume we have an oracle that can automatically turn an exploit deployed into a counter-exploit for the victim or potential victims with similar contracts.
As shown in \autoref{fig:timediff}, 50 out of 54 documented attacks have exploits deployed at least one second before the attacks. With such an oracle, \strategyone can rescue more than \$115M worth of assets from these 50 attacks.
Additionally, as shown in \autoref{fig:loss1}, we observe that the \strategytwo leveraging such an oracle can rescue more than \$246M worth of assets.

\subsection{Exploit Synthesis}

Unlike previous work, which synthesizes the counter-exploit after observing the full attack, \sys synthesizes the counter-exploit from the exploit contracts (\ie contracts deployed by attackers that are later triggered to conduct the attack) deployed before the attack transaction is sent. The challenge here is that the exploit contract commonly misses details about the attack. We use the Onyx Protocol hack~\cite{onyx} as an example:

\begin{figure}
    \centering
    \begin{lstlisting}
function 0xcb0d9b88(uint256 v0, bytes v1) public { 
    ...
    require(msg.sender == owner);
    require(tx.origin == msg.sender);
    require(0x60b0a6.... == keccak256(tx.origin));
    ...
    addr.flashloan(this, s19, v0, v0, 0);
    ...
    ret, res = stringToAddress(v1);
    require(owner == res);
    ...
}
\end{lstlisting}
    \caption{Decompiled code of exploit targeting Onyx Protocol}
    \label{fig:onyx}
\end{figure}

In the exploit contract, there are seven callable functions. The specific function that the attacker leveraged in the exploit contract is \verb|0xcb0d9b88|, as shown in \autoref{fig:onyx}. To use this function before seeing the attack transaction, we need to guess three input arguments: the sender, \verb|v0|, and \verb|v1|. The sender can be easily computed by trying all constants found in the contract storage and the code of the exploit contract.  However, finding suitable values for \verb|v0| and \verb|v1| is non-trivial: \verb|v0| controls the amount of flashloan borrowed; \verb|v1| is first converted to a string by \verb|stringToAddress| and then compared with the owner. \verb|v0| alone has $2^{256}$ possibilities, which cannot be brute-forced in a reasonable amount of time.   

Another example is the Grok attack, which is a price manipulation attack. After the initial attack, hundreds of victims remained vulnerable to the same attack. The first attacker used the following exploit contract. 

\begin{lstlisting}
function attack() public { 
    ...
    lp.buyToken(A)
    token.transferFrom(address(this), token, B)
    lp.buyToken(C)
    lp.sellToken(D)
    ...
}
\end{lstlisting}

In the exploit, the attacker hardcoded four \verb|uint256| constants denoted by \verb|A, B, C, D|, which only work for the initial victims. The values are directly correlated to the success of the price manipulation. To reuse the exploit for a different victim, we must find the values for \verb|A, B, C, D| specific to the victim. 

\vspace{5px}
\noindent\fbox{\begin{minipage}{23.5em}
\textbf{Challenge}: Turning an existing exploit into a defense exploit 
requires synthesizing complex patches or new code. 
\end{minipage}}
\vspace{5px}

\section{Methodology}

\begin{algorithm}[t!]
\begin{minipage}{\linewidth}
\caption{Overall Workflow}
\SetKwData{t}{t}
\SetKwData{Empty}{empty}
\SetKwData{Is}{is}
\SetKwData{to}{t'}
\SetKwData{Programs}{$\mathtt{Programs}$}
\SetKwData{Program}{$\mathtt{program}$}
\SetKwData{Profit}{$\mathtt{profit}$}
\SetKwData{E}{$E$}
\SetKwData{In}{in}
\SetKwData{Input}{Input}
\SetKwFunction{Extend}{extend}
\SetKwFunction{ExploitCloning}{ExploitCloning}
\SetKwFunction{Rewrite}{Rewrite}
\SetKwFunction{Maximize}{HybridFuzzing}
\SetKwFunction{Send}{Send}
\Input: $T$\;
\everypar={\nl}

$\Programs \gets$ \ExploitCloning($T$)

\For{$\Program \in \Programs$} {
$\Program' \gets$ \Rewrite{\Program}

\Profit, h $\gets$ \Maximize{\Program'}

\If{\Profit $>$ 0}{
\Send{\Program', h}
}
}
\end{minipage}
\label{algorithm:overview}

\end{algorithm}

\subsection{Threat Model}

Our approach targets attacks that exploit vulnerabilities in on-chain smart contracts.
Attacks stemming from other causes, including private key leaks, are out of the scope of this work.
Attacks launched by privileged parties themselves, such as rug pull scams, are also beyond our scope.
As mentioned in \autoref{motivation-atk-frontrunning}, the prevalence of private transactions makes attack transactions invisible to the public.
Therefore, unlike existing approaches~\cite{yeim,xue2022preventing,qin2023blockchain}, our approach does not assume the availability of an attacking transaction before the block broadcast. 

\subsection{Overview of \sys}

For every public transaction in the mempool and private transaction seen after block broadcast, \sys analyzes every contract created in the transaction (\strategyone) and the transaction (\strategytwo). 
\sys involves four stages: \verb|ExploitCloning|, \verb|Rewrite|, \verb|HybridFuzzing|, and \verb|Send|. A high-level overview of the workflow is shown in Algorithm~1. \verb|ExploitCloning| (\autoref{sec:exploit-cloning1}, \autoref{sec:exploit-cloning2}) yields a set of programs with holes (unfilled constants in the program) to be repaired. During \verb|ExploitCloning|, \strategyone takes a contract to generate a set of programs that explores all paths of that contract, with holes as inputs of that contract. Attack backrunning instead takes a transaction and generates a set of programs that swap the called address (potential victims) with new victims and make call inputs sent from the exploit contract to the new victims as holes. \verb|Rewrite| (\autoref{sec:rewrite}) takes a program and uses a set of pre-defined rules to eliminate holes (e.g., flashloan amount which can be calculated from the execution of the new programs.) Then, \verb|HybridFuzzing| (\autoref{sec:fuzzing}) takes the modified program and attempts to maximize the profit received by our account during the program's execution by trying different values for the unfilled holes. If a profitable execution is found, \sys sends it to our block builder. The block builder finds the most profitable execution and converts it to a transaction. The block builder then builds the block with that transaction and continuously bids to the validators to surpass others' bids for block commitment.

\subsection{Generating Exploits for \strategyone}
\label{sec:exploit-cloning1}

\begin{algorithm}[t!]
\label{alg:ec1}
\begin{minipage}{\linewidth}
\caption{Exploit Cloning for \strategyonecapitalized.}
\SetKwData{Return}{Return}
\SetKwData{JUMPI}{$\mathtt{JUMPI}$}
\SetKwData{JUMPN}{$\mathtt{JUMPN}$}
\SetKwData{i}{$\mathtt{i}$}
\SetKwData{trace}{$\mathtt{trace}$}
\SetKwData{pcs}{$\mathtt{pc}$}
\SetKwData{StopInstructions}{$\mathtt{StopInstructions}$}
\SetKwData{Contract}{$\mathtt{Contract}$}
\SetKwData{Default}{$\mathtt{Default}$}
\SetKwData{Input}{Input}
\SetKwData{Programs}{$\mathtt{Programs}$}
\SetKwData{Program}{$\mathtt{program}$}
\SetKwData{Func}{$\mathtt{func}$}
\SetKwData{Var}{$\mathtt{Var}$}
\SetKwData{Function}{Function}
\SetKwData{nth}{$\mathtt{nth}$}
\SetKwData{Stack}{$\mathtt{Frontiers}$}
\SetKwFunction{ExtractFuncs}{ExtractFuncs}
\SetKwFunction{ExploitCloning}{ExploitCloning}
\SetKwFunction{Push}{push}
\SetKwFunction{Append}{append}
\SetKwFunction{StartCall}{StartCall}
\SetKwFunction{ForceExecution}{Force}
\SetKwFunction{Pop}{pop}
\SetKwFunction{Eval}{Step}
\SetKwFunction{Add}{add}
\SetKwFunction{Instr}{Instr}
\SetKwFunction{epc}{entry\_pc}
\SetKwFunction{StepWithFlip}{StepWithFlip}
\SetKwFunction{instrsAt}{inst\_at}
\SetKwFunction{Remove}{remove}
\small
\Function: \ExploitCloning

\Input: \Contract\

\Return: \Programs $\gets \emptyset$
\everypar={\nl}

\For{\Func $\in$ \ExtractFuncs{\Contract}} {
    $\Stack \gets \{(\epc{\Func}, [])\}$
    
    \While{$(\pcs, \trace) \in {\Stack}$}{

    $\Stack = \Stack \setminus (\pcs, \trace)$

    $\i \gets \instrsAt{\Func, \pcs}$

    \While{$\i \notin \StopInstructions$}{
    
        $\trace \gets \trace + \pcs$
    
        \If{\i = \JUMPI}{

            \pcs' $\gets$ \StepWithFlip{\Func, \pcs}
            
            \Stack $\gets$ \Stack + (\pcs', \trace)
        }
        
        \pcs $\gets$ \Eval{\Func, \pcs}

    $\i \gets \instrsAt{\Func, \pcs}$
    }

    \Programs $\gets$ \Programs + ($\Func$, \trace)
    }
}
\end{minipage}
\end{algorithm}

In \strategyone, \sys analyzes a contract by first extracting all functions in the contract through decompilation. Each function has a set of arguments whose types \sys infers during decompilation. The programs, that \sys returns in exploit cloning are a set of functions with arguments as holes. 

Attacker commonly has checks in their contracts (e.g., authentication). We recognize that if the contract is an exploit, then there must be an execution path of one of the functions that leads to an attack.
With this insight, we must collect every path (including infeasible paths) in the contract.
%
Static analysis is a common way to explore all feasible paths and some infeasible paths in a contract without generating inputs. \sys does not use static analysis because: 1) exploit contracts often include external calls and callbacks, which static analysis cannot adequately handle at the bytecode-level\cite{slither,panoramix};
2) the targets of smart contract conditional jumps are mostly dynamically calculated, making path inference complex and time-consuming\cite{gigahorse}.  

We propose an unsound but practical dynamic approach to collect all feasible paths.  To explore all paths in a function precisely, we need to generate inputs for those paths using fuzzing or concolic/symbolic~\cite{dart,cute,klee,symcc} execution, which is infeasible because of their high computation cost. Therefore, we force 
exploring both branches of each conditional statement while running the function with default inputs (e.g., 0 for \verb|uint256|).  Even though exploration of both branches of each conditional jump may change the semantics of the exploit contract, \sys tries every possible branching combination (\ie path), and at least one combination would be semantically valid. 

%
%
The dynamic approach is very similar to generational concolic execution~\cite{GodefroidSAGE}; however, it does not try to check the feasibility of each path by generating inputs.  Therefore, the analysis is extremely fast and can handle external calls and callbacks. In real-world experiments, dynamic analysis consistently generates a concise set of paths, always including those taken by attackers. 

%
%
%
%
%
The pseudocode is shown in Algorithm~2. In the algorithm, we use \verb|+| to denote adding an element to a list, set, or map.  The algorithm maintains a set, called $\mathtt{Frontiers}$, of partial paths to be explored further by the algorithm.  A partial path is a pair whose first element is the PC that needs to be explored next and a sequence of program counters (\ie a partial trace) corresponding to conditional jumps already explored by the execution. At the start of the analysis of every function, $\mathtt{Frontiers}$ is initialized to a pair containing the entry program counter (PC) of that function, and an empty trace (line 2). Then, while there are partial paths in the set, \sys executes the function with default inputs (lines 5 - 12) and records every PC encountered during execution in the trace. 
For every occurrence of $\mathtt{JUMPI}$ during execution, \sys force executes to the other branch (line 9) and adds the resulting state to the $\mathtt{Frontiers}$ (line 10) for future forced exploration. 
Once the execution is finished, the current execution with the trace is added to the returned programs (line 13).

\subsection{Generating Exploits for \strategytwo}
\label{sec:exploit-cloning2}
\begin{algorithm*}[t!]
\caption{Exploit Cloning for \strategytwocapitalized}
\SetKwData{Tx}{$\mathtt{Transaction}$}
\SetKwData{r}{$\mathtt{r}$}
\SetKwData{JUMPN}{JUMPN}
\SetKwData{i}{$\mathtt{i}$}
\SetKwData{trace}{$\mathtt{transaction'}$}
\SetKwData{pc}{$\mathtt{pc}$}
\SetKwData{StopInstructions}{$\mathtt{StopInstructions}$}
\SetKwData{Contract}{$\mathtt{Contract}$}
\SetKwData{Input}{Input}
\SetKwData{Trace}{$\mathtt{transaction'}$}
\SetKwData{Action}{$\mathtt{action}$}
\SetKwData{Addr}{$\mathtt{addr}$}
\SetKwData{Args}{$\mathtt{arg}$}
\SetKwData{Stack}{$\mathtt{Stack}$}
\SetKwFunction{Reconstruct}{Reconstruct}
\SetKwFunction{ReplaceAddr}{ReplaceAddr}
\SetKwFunction{Add}{add}
\SetKwData{ReplacedAddr}{$\mathtt{Replacers}$}
\SetKwFunction{NewAddr}{NewAddr}
\SetKwFunction{Push}{push}
\SetKwFunction{Extend}{extend}
\SetKwFunction{Pop}{pop}
\SetKwFunction{IsConstant}{IsConstant}
\SetKwFunction{Filtered}{filter}
\SetKwData{Function}{Function}
\SetKwData{Return}{Return}
\SetKwFunction{Replace}{Overwrite}
\SetKwData{Traces}{$\mathtt{ModifiedTransactions}$}
\SetKwFunction{InvolvedAddrs}{addrs\_used}
\SetKwFunction{Map}{Map}
\SetKwFunction{EmptyMap}{EmptyMap}
\SetKwFunction{Values}{values}
\SetKwFunction{ArgsOf}{args\_of}
\SetKwFunction{Trait}{Traits}
\SetKwData{Programs}{$\mathtt{Programs}$}
\SetKwData{ReplacedAddrSingle}{$\mathtt{SimilarAddrs}$}
\SetKwData{SimilarAddr}{$\mathtt{addr'}$}
\SetKwData{Replacer}{$\mathtt{replacer}$}
\SetKwData{Conforms}{$\mathtt{Conforms}$}
\SetKwData{tx}{$\mathtt{transaction}$}
\SetKwData{Traits}{$\mathtt{traits}$}
\SetKwData{Holes}{$\mathtt{Holes}$}

\small
\Function: \ExploitCloning

\Input: \Tx\

\Return: \Programs
\everypar={\nl}

\Trace $\gets$ \Reconstruct{\Tx}

\ReplacedAddr $\gets \{\EmptyMap{}\}$

\For{\Action $\in$ \Trace} {
\ReplacedAddrSingle $\gets \{\}$ 

\For{\Addr $\in$ \InvolvedAddrs{\Action}} {

    \ReplacedAddrSingle[\Addr] $\gets \{\Addr' \mid \Trait(\Addr') = \Trait(\Addr) \text{ for all } \Addr' \text{ on chain}\}$
}


\For{\Addr $\in$ \InvolvedAddrs{\Action}}{
$\ReplacedAddr' \gets \{\}$    

    \For{\Replacer $\in$ \ReplacedAddr $\wedge$ \SimilarAddr $\in$ \ReplacedAddrSingle} {
        
            
            \If{$\forall (k \mapsto v) \in \Replacer \wedge r(\Addr, k) = r(\Addr', v)$} {
                $\Replacer \gets \Replacer + (\Addr \mapsto \SimilarAddr)$
            
            $\ReplacedAddr' \gets \ReplacedAddr' + \Replacer$
            }

    }

    $\ReplacedAddr \gets \ReplacedAddr'$
}}

$\Traces \gets \{\Replacer(\Trace) | \Replacer \in \ReplacedAddr\}$

$\Holes \gets \{\}$

\For{\Action $\in$ \Trace} {
    \For{\Args $\in$ \ArgsOf{\Action}}{
        \If{\IsConstant{\Args}}{
            $\Holes \gets$ $\Holes$+ (\Action, \Args)
        }
    }
}
$\Programs \gets \{({\tx}, \Holes)|\tx \in \Traces\}$
\end{algorithm*}

In \strategytwo, an attack transaction is first reconstructed such that \sys can gain the same profit as the attacker originally gained by conducting the same attack. Attack reconstruction has been well-explored in previous work of attack frontrunning \cite{yeim,girlfriend}. 
The reconstruction process first transforms an attack transaction into a sequence of actions the attackers took, such as external calls, token liquidations, flashloans, etc., using pattern matching. 
Then, to redirect the attack's profit from the attacker to us, the reconstruction process replaces all occurrences of exploit contracts and attacker addresses with addresses we can control. 






Each attack has a set of victims, which are the addresses that lose funds in the attack and the addresses called by these addresses. After the reconstruction, \sys swaps the original victim set to other potential victim sets to derive a set of new attack transactions. 
New potential victims for every original victim in the victim set are found by matching addresses on the chain with similar traits as the original victim.
For instance, if a contract address has the same trait (e.g., same bytecode or function signatures) as one of the original victims, then it can be considered as a new potential victim that can swap that original victim. 
%
%

As depicted in Algorithm~3, to identify relevant new victims for an attack, the algorithm scans each address \(\mathtt{addr}\) involved in the attack (lines 4-6). For each \(\mathtt{addr}\), \(\{\mathtt{addr}' \mid \text{Traits}(\mathtt{addr}') = \text{Traits}(\mathtt{addr})  \text{ for all } \mathtt{addr}' \text{ on chain}\}\) forms a set of potential victims, characterized by traits equivalent to those of \(\mathtt{addr}\) (line 6).

Despite the precision of trait definitions, the number of potential new victims can be exceedingly large (\(>10^8\)), making the number of potential new victim sets even larger. An important insight is that new victims in each potential victim set must exhibit similar correlations to each other as their counterparts in the original victim set. For example, in a victim set comprising $\mathtt{(token, lp)}$, where $\mathtt{lp}$ is the liquidity pool of the token $\mathtt{token}$. As liquidity pool contracts typically share identical function selectors and bytecode, \(\mathtt{SimilarAddrs}[\mathtt{lp}]\) contains over 5 million $\mathtt{lp'}$ on Ethereum such that $\texttt{Traits}(\mathtt{lp'}) = \texttt{Traits}(\mathtt{lp})$.

Thus, for every set in $\{\mathtt{(token', lp')} \mid \mathtt{token'} \in \mathtt{SimilarAddrs}[\mathtt{token}] \wedge  \mathtt{lp'} \in \mathtt{SimilarAddrs}[\mathtt{lp}]\}$
identified by trait matching, \sys must additionally filter \(\mathtt{lp'}\) instances to retain those which specifically serve as the liquidity pool for \(\mathtt{token'}\) rather than unrelated tokens. 
To formally define the correlation filter, we introduce \(r(\mathtt{addr}, \mathtt{addr'})\), 
a set of manually defined rules returning the correlations (represented as a set of relationships) between any two addresses. \sys eliminates victim sets where every address combination 
\((\mathtt{addr'}, v)\) does not satisfy \(r(\mathtt{addr}, k) = r(\mathtt{addr'}, v)\), with \(\mathtt{addr}\) and \(k\) being the original counterparts of \(\mathtt{addr'}\) and \(v\) respectively (line 10).

Using the previous example, suppose \(\mathtt{SimilarAddrs}[\mathtt{token}]\) contains a single address \(\mathtt{token'}\). At Line 9, since the replacer is initially empty, \sys takes the true condition branch and updates the replacer to \(\{\mathtt{token} \mapsto \mathtt{token'}\}\) (Lines 10-12). Subsequently, the algorithm processes each \(\mathtt{lp'} \in \mathtt{SimilarAddrs}[\mathtt{lp}]\). For each \(\mathtt{lp'}\), the mapping \(\{\mathtt{lp} \mapsto \mathtt{lp'}\}\) is appended only if \(r(\mathtt{lp}, \mathtt{token}) = r(\mathtt{lp'}, \mathtt{token'}) = \{\text{LP}\}\) (Line 10). Finally, the \verb|Replacers| set comprises a set of mappings in the form \((\mathtt{token} \mapsto \mathtt{token'}, \mathtt{lp} \mapsto \mathtt{lp'})\), where \(\mathtt{lp'}\) is the liquidity pool of \(\mathtt{token'}\).

Additionally, each external calls in the reconstructed trace have arguments, which are either constants or returns of the previous call. Arguments may need to be modified when applying the attack on different victims. Thus, if the argument is a constant, \sys considers it to be a hole that needs to be filled with a new value (lines 14-19). 

\subsection{Rule-based Exploit Rewrite}
\label{sec:rewrite}
We discovered that some holes could be removed by applying a set of manually crafted rewriting rules. Reducing the number of holes can reduce complexity in the next step as fewer holes need to be filled.
We utilize three practical exploit rewrite rules to reduce the number of holes.

\vspace{5px}
\noindent\textbf{Flashloan} Attackers may borrow flashloan to conduct the attack. The exact amount of flashloan needed is the amount spent in the subsequent external calls.
Thus, \sys fills holes passed as the amount in flashloan calls with such values.
Some attacks may need more funds than those in a single flashloan provider, and they would need to borrow flashloan from multiple providers. In this situation, \sys starts by borrowing from providers with the lowest fee until reaching the amount needed. 

\vspace{5px}
\noindent\textbf{Approval Amount} Some attacks attempt to drain tokens approved by accounts to the victim contract. \sys replaces holes that are passed as the amount used in calls transferring approved tokens with results of \verb|allowance(spender, owner)|, which returns the tokens approved.    

\vspace{5px}
\noindent\textbf{Swap} Attackers may swap a token to another token during the attack. To perform swap, the attacker calls \verb|swap(amount0Out, amount1Out, ...)| for Uniswap V2 liquidity pool, and calls function \verb|swap(..., direction, amount, price, ...)|  for Uniswap V3 pool. \verb|amount0Out|, \verb|amount1Out|, \verb|price| before and after the attack transaction can be different and need to be recalculated based on liquidity in the pool.
For holes related to those arguments, \sys replaces them with proper swap calculation logic.

\subsection{Hybrid Fuzzing-based Exploit Repairing}
\label{sec:fuzzing}

To fill the remaining holes, \sys uses hybrid fuzzing, which combines fuzzing with symbolic execution.  It might appear that symbolic execution can effectively fill the holes and find profitable executions.
However, in the real world, symbolic execution can generate gigabytes of constraints that take days to solve. Even worse, DeFi projects such as Uniswap V3 have complex loop invariants and conduct square roots and logarithmic operations, further hindering constraint solving.
Another observation is that the programs with holes may be easier to solve if we can simply prepend or append more calls to them. For instance, a revert may be caused by a lack of a specific token in the victim contract. We can easily resolve this revert by appending flash-loan and swapping transactions to get that token.
With these observations, we use coverage and dataflow-guided stateful fuzzing~\cite{ityfuzz} with concolic execution~\cite{concolicfuzzsc} to find valid inputs. The concolic execution module and fuzzing module share the same corpus (\,  i.e., a set of interesting inputs found that are later retrieved for mutations and deriving new inputs), and \sys run both modules each on multiple threads concurrently. 
The fuzzer treats the programs with holes as potential callable functions and is encouraged to prepend and append more calls when filling the holes. 

As \sys needs to react fast in the real world, we introduce heuristics to reduce the input space to speed up the hybrid fuzzing process. 
First, \sys only explores inputs that can be decoded with respect to the inferred type (e.g., only attempt $[0, 2^{16}-1]$ for uint16).
Second, we extract all constants and state variables from all contracts called and use them in the initial corpus.
Lastly, we observe that static call returns and values in the EVM stack observed during execution may also be valid input arguments. \sys uses these values as hints for the mutators (e.g., the mutators may replace some bytes in the input with values from the hints). 

We also use the potential profit gained from execution to assign energy to each test case, maximizing the profit until the time limit of fuzzing is reached. 
The energy of a test case determines the computational power to be spent by the fuzzer on it.
A test case with higher energy will be used in fuzzing for a longer time, thus generating more new inputs.
Assuming $p$ is the profit, we assign each test case energy $e'(t)$ based on \autoref{equ:power-schedule}.
    \vspace{-10px}

\begin{equation}
\label{equ:power-schedule}
e'(t) = min(32 * e(t), 100 * log_2(max(2, p)))
\end{equation}
Here, $e(t)$ is the original power scheduler assigned energy. 
We first take the logarithmic value of $p$ to scale down the energy difference in test cases of large $p$.
To avoid assigning too much energy, we cap the new energy at 32 times the original energy.
We chose the coefficients because they empirically work best in the experiments.
With profit-guided power scheduling, \sys schedules the fuzzing power toward discovering inputs covering greater profits, therefore helping the fuzzer achieve its goal faster and more efficiently.

\section{Evaluation}

\subsection{Experiment Setup}
We have implemented a frontrunning bot based on STING~\cite{yeim} in 67K lines of code in Rust. Then, we implemented our \strategyone and \strategytwo techniques on top of it using 38K lines of Rust code. We leveraged \texttt{revm}\cite{revm} to simulate and trace transactions. We also reused taint analysis, fuzzing, and concolic execution modules in ItyFuzz~\cite{ityfuzz}. In addition, to identify similar contracts, we indexed 46TB of traces and contracts on Ethereum, Binance Smart Chain, Arbitrum, and Polygon. 

For back-testing (testing on past data), we created a simulated blockchain node, which replays each block using our collected trace, along with Blocknative mempool data. We impose that the time interval between the bots seeing the transaction and block broadcast is exactly the same as in the real world. We used an AWS spot instance cluster of 1024 cores with 16TB memory for hybrid fuzzing. For baseline, we compare \sys with our implementation of STING\cite{yeim}, the state-of-the-art frontrunning bot in academia. Additionally, we also compare with existing attack frontrun bots \cite{wellknowbots} on the chains. These bots can synthesize counter exploits with complex control flow and dataflow mutations in less than 100ms. 

For real-world settings, we deployed three bots (each running with 192 cores and 768 GB memory) connected to bloXroute BDN to process transactions on Ethereum, Polygon, Binance Smart Chain, and Base worldwide (to access public transactions sent in the region with low latency). Each bot is deployed along with a Reth\cite{reth} or Geth\cite{geth} full node.
%

\subsection{Performance on Past Attacks}
\subsubsection{Preemptive Hijack Performance}

We run \strategyone of \sys on exploit contracts collected from 54 initial attack transactions on Binance Smart Chain, Ethereum, Arbitrum, and Base, and \strategyone can generate a defense exploit for 38 of them. 
Of these 54 exploit contracts, 14 of them can be directly converted into defense exploits using exploit cloning. After exploit cloning, the maximum amount of holes in exploits is 26, and the subsequent steps need to identify a valid value for these holes. Rule-based rewrites can yield 8 additional valid exploits and reduce the number of holes for 6 projects. Finally, fuzzing-based repair generates an additional 16 exploits. In total, within 1.5 years since 2023, \strategyone can rescue \stratonebt, which is 15.6\% more than the funds stolen by the attackers and 387\% more than the funds rescued via the baseline technique. \sys can rescue even more than the attackers profited because the defense exploits generated by \sys use cheaper flashloan and holes yielding the most profit, etc.

\begin{figure}
    \centering
    \includegraphics[width=8cm]{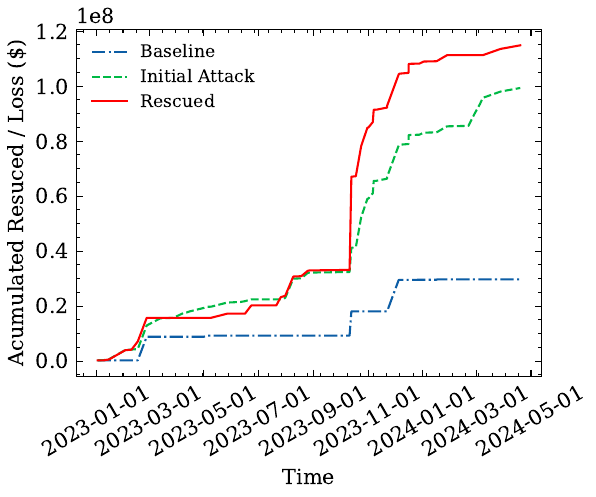}
    \caption{Preemptive hijack and attack frontrunning (baseline) rescued versus loss caused by initial attack transactions.}
    \vspace{-5px}
    
    \label{fig:rescued1}
\end{figure}

We show the performance of \strategyone on these exploits in ~\autoref{table:fr} and \autoref{fig:rescued1}.   
The exploit cloning mechanism can instantly generate a defense exploit for exploits targeting projects such as BarleyFi\cite{barley},  TransitFi\cite{transit}, and BEARNDAO\cite{bearn}.
Specifically, attack functions in these exploits either have no argument or have arguments that have no impact on the subsequent execution. These exploits can be simply turned into a \strategyone exploit by bypassing certain simple checks in the control flow.

Rule-based rewrites can eliminate and fill the holes in cases such as NFT Trader\cite{nftrade}. Specifically in NFT Trader, by updating the argument of the action used to drain victims' NFTs with the number of victims owned and approved to the vulnerable contract, rule-based rewrite can derive a functional exploit that drains the remaining NFTs.  
%
%
%

As discussed, \sys leverages fuzzing to fill the rest of the holes.
Yet, there are 12 cases where, even with hybrid fuzzing, \sys cannot fill the holes properly.
Although these cases are rare, we discuss them to understand \sys's ability better.
There are mainly two categories that \sys fails to handle.
First, \sys fails to handle attacks targeting projects such as Unizen\cite{unizen} and WOO\cite{woo} because the attacker conducted the attack in a single transaction. Specifically, they conducted the attack inside the constructor of the exploit contract, and once deployed, the attack was finished. Without the attack contract, \sys cannot conduct \strategyone. We further discuss such a weakness in \autoref{sec:discussions}. 
%
%

%
Additionally, for exploits targeting projects such as PawnFi, \sys cannot find valid hole values even after hybrid fuzzing. 
The reason is that these holes are hard to fill. Specifically for PawnFi\cite{pawnfi}, one hole is used by the exploit as the value needed to enter markets in a DeFi project.
In the experiment, we observe that filling the hole with values out of range $[2e23-1e18, 2e23 + 1e18]$ makes the full exploit revert.
Fuzzing is impractical to solve this constraint with limited computation resources and time. 
Meanwhile, due to the extremely complex constraints introduced by Compound, a liquidity pool used by the exploit, concolic execution aborts early before even generating constraints for such a hole.
\sys fails to generate defense exploits for holes involving extremely complex constraints.

 \begin{table*}[!ht]

    \centering
\small
\vspace{1cm}

\begin{tabular}{l|c|c|ccccc|cccccc}
\multirow{2}{*}{\textbf{Incident}} &
\multirow{2}{*}{\textbf{\textbf{Date}}} &
      \multicolumn{1}{|c|}{\textbf{Baseline}} &
      \multicolumn{5}{c|}{\textbf{\strategyonecapitalized}} &
      \multicolumn{6}{c}{\textbf{\strategytwocapitalized}} \\
                  &  & \textbf{R/I} & \textbf{EC}           & \textbf{H} & \textbf{RR}            & \textbf{H'} & \textbf{R/I} & \textbf{\#E} & \textbf{H}            & \textbf{R}            & \textbf{H'} & \textbf{R/T} & \textbf{R/B} \\
                   \hline
Hedgey & 2024-04-19 & 0\% & \xmark & 9 & \xmark & 1 & 99.86\% & 7 & 17 & \cmark & 16 & 90.85\% & 239.65\%     \\
PrismaFi & 2024-03-28 & 0\% & \cmark & - & - & - & 99.84\% & 3 & 1 & \xmark & 1 & 118.56\% & 149.44\%     \\
unizen & 2024-03-08 & 0\% & - & - & - & - & 0\% & 875 & 8 & \xmark & 2 & 58.65\% & 266.06\%     \\
WOO & 2024-03-05 & 0\% & - & - & - & - & 0\% & 1 & - & - & - & 0.00\% & 0.00\%       \\
Seneca & 2024-02-28 & 0\% & \xmark & 0 & \xmark & 0 & 0\% & 6 & 5 & \xmark & 3 & 37.96\% & 127.44\%     \\
DN\_404 & 2024-02-21 & 0\% & \xmark & 1 & \xmark & 1 & 0\% & 1 & 7 & \xmark & 2 & 0.27\% & inf          \\
BarleyFi & 2024-01-28 & 100\% & \cmark & - & - & - & 100.00\% & 2 & 8 & \xmark & 4 & 16.34\% & 174.90\%     \\
BasketDAO & 2024-01-17 & 0\% & \xmark & 1 & \xmark & 1 & 105.16\% & 10 & 1 & \cmark & - & 28.44\% & inf          \\
Socket & 2024-01-16 & 0\% & - & - & - & - & 0\% & 5.7K & - & - & - & 44.35\% & 225.26\%     \\
Radiant & 2024-01-02 & 0\% & \xmark & 3 & \xmark & 3 & 0\% & 1 & - & - & - & 0.00\% & 0.00\%       \\
TransitFi & 2023-12-19 & 100\% & \cmark & - & - & - & 100.00\% & 3.1K & 6 & \xmark & 5 & 589.88\% & inf          \\
NFT Trader & 2023-12-16 & 0\% & \xmark & 9 & \xmark & 1 & 18214\% & 8.2K & 8 & \xmark & 7 & 1276.2\% & 1283.3\%    \\
INS20 & 2023-12-28 & 0\% & \xmark & 1 & \cmark & - & 100.00\% & 21M & 1 & \cmark & - & 419.08\% & 420.35\%     \\
Floor NFT & 2023-12-16 & 0\% & \xmark & 0 & \xmark & 0 & 452.75\% & 784 & 1 & \cmark & - & 187.83\% & 321.02\%     \\
Elephant & 2023-12-06 & 0\% & \xmark & 1 & \xmark & 1 & 100.00\% & 1.4K & 13 & \xmark & 1 & 1.26\% & 4.11\%       \\
BEARNDAO & 2023-12-05 & 0\% & \cmark & - & - & - & 100.00\% & 2.9K & 10 & \xmark & 1 & 6.07\% & inf          \\
Kyberswap & 2023-11-22 & 0\% & - & - & - & - & 0\% & 1 & - & - & - & 0.00\% & 0.00\%       \\
Bot 0x8c2d & 2023-11-22 & 0\% & \xmark & 1 & \cmark & - & 94.33\% & 2 & 19 & \xmark & 18 & 0.00\% & inf          \\
Raft & 2023-11-10 & 0\% & \xmark & 2 & \xmark & 1 & 100.00\% & 1 & - & - & - & 0.00\% & 0.00\%       \\
Bot 0x05f0 & 2023-11-07 & 0\% & \xmark & 1 & \cmark & - & 100.06\% & 13 & 4 & \xmark & 1 & 36.91\% & 1878.4\%    \\
TheStandard & 2023-11-06 & 0\% & \cmark & - & - & - & 100.00\% & 49 & - & - & - & 0.00\% & 0.00\%       \\
Onyx & 2023-11-01 & 0\% & \xmark & 2 & \cmark & - & 100.12\% & 1 & - & - & - & 0.00\% & 0.00\%       \\
UniBot & 2023-10-31 & 0\% & \xmark & 0 & \xmark & 0 & 0\% & 1.8K & - & - & - & 1245.6\% & 5506.8\%    \\
Maestro & 2023-10-24 & 0\% & \cmark & - & - & - & 100.00\% & 37K & 3 & \cmark & - & 243.28\% & 252.75\%     \\
Hope.money & 2023-10-18 & 100\% & \xmark & 5 & \xmark & 3 & 0\% & 4 & - & - & - & 0.00\% & 0.00\%       \\
WiseLending & 2023-10-13 & 100\% & \cmark & - & - & - & 100.00\% & 1 & - & - & - & 0.00\% & 0.00\%       \\
BH Token & 2023-10-11 & 0\% & \xmark & 8 & \xmark & 8 & 0\% & 3 & 14 & \xmark & 1 & 3.69\% & inf          \\
Balancer & 2023-08-27 & 0\% & \xmark & 26 & \xmark & 26 & 0\% & 1 & - & - & - & 0.00\% & 0.00\%       \\
SVT & 2023-08-25 & 0\% & \xmark & 2 & \xmark & 2 & 100.00\% & 1 & 6 & \xmark & 4 & 0.48\% & inf          \\
Exactly & 2023-08-18 & 0\% & \xmark & 2 & \xmark & 2 & 100.00\% & 1 & 14 & \xmark & 13 & 24.59\% & inf          \\
EarningFarm & 2023-08-09 & 0\% & \xmark & 1 & \cmark & - & 100.00\% & 1 & - & - & - & 0.00\% & 0.00\%       \\
LeetSwap & 2023-07-31 & 0\% & \cmark & - & - & - & 100.00\% & 7 & \cmark & - & - & 63.98\% & 97.84\%      \\
Curve & 2023-07-30 & 100\% & \cmark & - & - & - & 100.00\% & 121 & 9 & \xmark & 3 & 42.36\% & 52.36\%      \\
Carson & 2023-07-26 & 0\% & \xmark & 4 & \xmark & 4 & 100.00\% & 1 & 17 & \xmark & 1 & 27.50\% & 83.62\%      \\
ConicFi & 2023-07-21 & 0\% & \xmark & 4 & \xmark & 4 & 101.20\% & 1 & - & - & - & 0.00\% & 0.00\%       \\
Shido & 2023-06-23 & 0\% & \xmark & 2 & \xmark & 2 & 0\% & 28 & - & - & - & 0.00\% & 0.00\%       \\
PawnFi & 2023-06-16 & 0\% & \xmark & 2 & \xmark & 2 & 0\% & 2 & - & - & - & 0.00\% & 0.00\%       \\
Sturdy & 2023-06-11 & 0\% & \xmark & 4 & \xmark & 4 & 100.00\% & 6 & - & - & - & 0.00\% & 0.00\%       \\
BabyDoge & 2023-05-27 & 0\% & \xmark & 1 & \xmark & 1 & 100.00\% & 1 & 20 & \xmark & 8 & 10.91\% & inf          \\
SNK & 2023-05-09 & 0\% & \xmark & 4 & \xmark & 1 & 100.00\% & 1 & 6 & \xmark & 4 & 108.1\% & 118.83\%     \\
DEI & 2023-05-05 & 0\% & \xmark & 1 & \cmark & - & 100.00\% & 2 & 3 & \cmark & - & 0.06\% & inf          \\
LevelFi & 2023-05-01 & 0\% & \xmark & 1 & \xmark & 1 & 100.00\% & 1 & - & - & - & 0.00\% & 0.00\%       \\
HundredFi & 2023-04-15 & 0\% & \xmark & 2 & \xmark & 1 & 100.00\% & 2 & - & - & - & 0.00\% & 0.00\%       \\
yearnFi & 2023-04-12 & 0\% & \xmark & 1 & \xmark & 1 & 100.00\% & 1 & - & - & - & 0.00\% & 0.00\%       \\
Sushiswap & 2023-04-08 & 0\% & \xmark & 8 & \cmark & - & 1615.4\% & 154 & 1 & \cmark & - & 92.32\% & 97.81\%      \\
SafeMoon & 2023-03-28 & 100\% & \xmark & 2 & \xmark & 2 & 99.99\% & 14 & 10 & \xmark & 2 & 0.13\% & inf          \\
EulerFi & 2023-03-13 & 100\% & \cmark & - & - & - & 385.13\% & 3 & - & - & - & 97.06\% & 101.54\%     \\
swapX & 2023-02-26 & 0\% & \xmark & 16 & \xmark & 16 & 0\% & 1 & - & - & - & 0.00\% & 0.00\%       \\
Dexible & 2023-02-16 & 0\% & \cmark & - & - & - & 100.00\% & 2 & - & - & - & 3.60\% & inf          \\
dForce & 2023-02-09 & 0\% & \xmark & 1 & \cmark & - & 100.00\% & 155 & - & - & - & 0.00\% & 0.00\%       \\
Orion & 2023-02-02 & 0\% & \cmark & - & - & - & 100.00\% & 2 & 5 & \xmark & 2 & 94.83\% & 101.28\%     \\
Midas & 2023-01-15 & 0\% & \cmark & - & - & - & 100.00\% & 8.9K & - & - & - & 0.00\% & 0.00\%       \\
BRA Token & 2023-01-09 & 0\% & \cmark & - & - & - & 100.00\% & 1 & 13 & \xmark & 1 & 57.25\% & 332.09\%     \\
GDS Token & 2023-01-02 & 0\% & \cmark & - & - & - & 100.00\% & 2 & 9 & \xmark & 4 & 140.3\% & inf \\

\end{tabular}

    \caption{\strategyonecapitalized and \strategytwocapitalized Performance. \textbf{EC}, \textbf{RR} are whether an exploit can be generated directly by exploit cloning and rewrite (if there are holes, fill with default values), respectively. \textbf{H} and \textbf{H'} are the minimum \# holes in the exploit before and after rewrite. \textbf{\#E} shows the amount of programs with holes can be generated by exploit cloning. \textbf{R/I, R/B, R/T} is the amount rescued vs loss from initial attack, loss from subsequent attack, and total loss respectively.}
     \label{table:fr}

\end{table*}

\begin{figure}
    \centering
    \includegraphics[width=8cm]{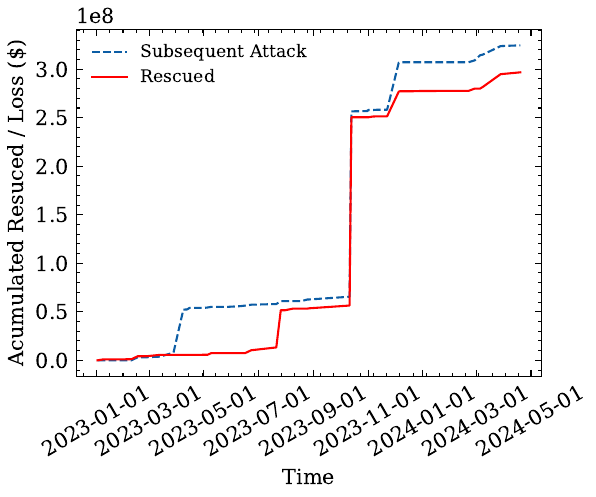}
    \caption{Attack backrunning rescued versus loss caused by subsequent attack transactions.}    
    \label{fig:rescued2}
    
\end{figure}


    

\subsubsection{Attack Backrunning Performance}
%
We used \sys to automatically backrun the initial attacks for all 54 initial attack transactions.
\sys generated at least one backrun exploit for each of the 33 projects with more than \$1K profits.
These generated exploits can gain \strattwobt in profits, rescuing 91.1\% assets from copycat or subsequent attacks. 
We show the performance of \strategytwo on these exploits in ~\autoref{table:fr} and \autoref{fig:rescued2}.

Specifically, in attacks targeting EulerFi \cite{euler}, \sys can rescue more than \$160M.
%
%
The exploit can be directly reused by replacing the victim set with any one of the pools of Euler Finance, its respective assets, or flash loan providers that can provide that asset.

In cases such as Socket\cite{socket} and Maestro\cite{maestro}, the exploits target victims who have approved funds to be spent by a vulnerable contract. \sys can easily recognize the approval relation between victims and the vulnerable contract during the exploit cloning phase. Then, the exploit cloning phase finds all other victims that have a similar relation to the contract. Replacing the original victims with new victims can easily generate a backrun exploit to extract all funds from those victims. 

After exploit cloning, \sys may generate backrun exploits with holes.
%
%
%
Rule-based rewrites are helpful as they can significantly reduce the number of holes.
Specifically, for the backrunning exploit of BEARNDAO\cite{bearn}, rule-based rewrite can reduce the number of holes from 10 to 1 by filling the holes in actions conducting flashloan, repayment, and Uniswap V3 swap with proper calculation logics. 
%

Finally, valid backrun exploits can be generated for most projects except projects such as Kyberswap\cite{kyber} in fuzzing-based repair.
The exploit of the initial attack on Kyberswap uses hardcoded four unique numbers to trigger an intricate rounding error in the victims.
Finding such four numbers is non-trivial; even with manually crafted invariant tests designed for this rounding error issue, it takes more than 30 seconds to find the desired test cases.
In the full exploit, finding these four numbers accurately, especially making the exploit profitable after leveraging the rounding error, is impractical with limited computation resources and time.
Yet, we suggest that human or large language model guidance can significantly help in this case. By localizing the rounding error with expert insights, one can craft an invariant test to speed up the process of identifying the values. 

\subsection{Performance in the Real-world}

We ran \sys in the real world in December 2023 and May 2024 to understand how the two strategies (\ie, \strategyone and \strategytwo) perform in terms of generating exploits and rescuing funds. \autoref{sec:ethical} discusses the details of the procedures we adopted.  

\sys successfully leveraged \stratonerescuedamt \strategyone opportunities on Base, Binance Smart Chain, and Ethereum, rescuing \stratonerescued automatically.
%
%
%
%
%
\sys failed to generate an exploit for one remaining opportunity, which is the Magic Internet Money attack\cite{mim}.
No existing bot managed to leverage this opportunity either.
However, our further investigation showed that such a failure happened because of a bug in \sys.  
%
After fixing the issue, we replayed the attack on our test chain and verified that \sys could successfully rescue the funds.

For \strategytwo, \sys successfully generated backrunning exploits for 18 attacks and rescued \$620K funds from more than 8.1K victims automatically. After tweaking values and inserting external calls to the \sys generated exploits manually, we rescued \$7.2M funds in two additional attacks. \sys ignored or filtered the additional 17 opportunities due to the limitation of computation resources. Future research can introduce additional optimization techniques.

\subsection{Computation Resource Cost}
\begin{figure}
    \centering
    \includegraphics[width=8cm]{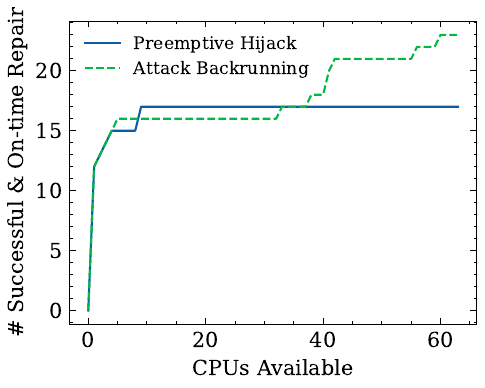}
    \caption{Average amount of successful and on-time fuzzing-based repair versus available CPUs.}
    \label{fig:cpus}
    
\end{figure}

In our implementation of \sys, exploit cloning and rule-based rewrite can finish in less than 300ms on average on a single core. Sending and triggering the generated exploits takes less than 10ms in real-world scenarios. 
For fuzzing-based repair, we observe that the time overhead can be reduced by parallelizing fuzzing with more CPU cores. 
For every \strategyone and \strategytwo attempts, we demonstrate how many cores are needed for each fuzzing-based repairs such that \sys can generate the defense exploit before attackers conduct the attack in \autoref{fig:cpus}. Among 40 attempts, 24 attempts can finish on time with only one core, and 30 attempts can do so with four cores. 
Yet, it is possible that by replacing the fuzzing engine used under the hood~\cite{smartest,smartian,wu2024yet,gpufuzz}, fuzzing-based repair can have even lower timing overhead with less computation resources. We leave this to future work.  


    
\section{Discussions}\label{sec:discussions}

\sys is designed to be a best-effort last-line solution for protecting smart contracts from attacks. That said, there are certain ways that attackers can bypass \sys. In the following sections, we describe potential methods to bypass \strategyone and \strategytwo. 

\subsection{Weakness of \strategyone}
\textbf{Attack from Launchpads.} We recognize that \sys would fail in the case when the exploit is in the form of launchpads. With the launchpad, the attack can later send private attack transactions that provide all the call information to the launchpads and finish the attack. An example of the launchpad is given below.

\begin{lstlisting}
function aggregate(Call[] memory calls)  {
    for(uint256 i = 0; i < calls.length; i++) {
        calls[i].target.call(calls[i].callData);
    }
}
\end{lstlisting}

In this launchpad, the attacker can call \verb|aggregate()| to conduct any types of attack by providing a sequence of call data. Yet, the assumption for \strategyone is that before the attack happens, \sys can gain leads on the attack, knowing the potential exploit sequences and traits of the victims, which in this case is not available. 
Unless one can see the private transaction, there is no possible solution to predict an attack launched by this method, regardless of how advanced the technique is. If one can predict it, then one can conduct the attack even without any knowledge about the attack, which is out-of-scope of this research. 
However, very few human attackers choose to launch attacks using launchpads because using them consumes significantly more gas than using conventional exploits. 
Deriving the input for launchpads is also extremely complex, error-prone, and time-consuming. All known attacks using launchpads are conducted by attack frontrunning bots or fuzzing bots.

\vspace{10px}
\noindent\textbf{Atomic and Bundled Attacks.} \sys would also fail when the exploit deployment and the attack transaction happen in the same block, both sent as private transactions. However, we have not yet observed any attacker has leveraged such a method. Another method is to conduct the attack in the exploit constructor. By doing so, the attack happens during exploit deployment. If the exploit is deployed with private transactions, there is again no possible solution to defend against it. In the last 1.5 years, we only observed 4 attack transactions leveraged constructors to conduct attacks. These two methods are rarely used by attackers due to their complexity. 

\subsection{Weakness of \strategytwo}

\noindent\textbf{Exploit Extracting All Funds.} \sys can leverage no backrunning opportunities if the exploit manages to steal all funds from all deployments of the projects. However, in the real world, this is very rare. An attack can use a high amount of gas. Launching the same attack targeting different victims takes a huge amount of gas, which would surpass the limit on the total gas of the block. Attackers typically run exploits on multiple blocks to attack different deployments to avoid this issue. Yet, after the block is broadcasted, the initial attack would be seen by \sys, and a backrun exploit can be generated in hundreds of milliseconds, blocking the subsequent attacks. Another scenario is that the attack would happen on multiple chains simultaneously. However, it is impossible for different chains to have the same block broadcast time. A few hundred milliseconds are enough for \sys to capture the backrunning opportunities. 

\vspace{10px}
\noindent\textbf{Adaptive Obfuscation and Hardcoded Values.} While \sys employs effective repairing approaches to generate exploits in most cases, the system's reliance on the initial exploit used leaves it vulnerable. Attackers could design exploits that circumvent \sys by including numerous extraneous external calls, forcing \sys to produce many unnecessary holes needing time-consuming fuzzing. Additionally, exploits using hardcoded values tailored to specific victims would compel \sys to regenerate fitting values when deployed against different targets. Though \sys can typically determine appropriate hardcoded values rapidly, adversaries could potentially leverage these obfuscation techniques to bypass \sys’s exploit generation defenses. Further research into hardening \sys against these attack avenues could make the system more robust.

\section{Related Works}

\textbf{Program Repair.} Different methods for program repair\cite{repair1,repair2,repair3} have been proposed in the last decades. For smart contracts, recent research works leverage search-based software engineering\cite{search0r}, reinforcement learning\cite{rl1r}, rule-based rewrite\cite{rule0r,rule1r,rule2r}, and semantic-based rewrite\cite{semanticr} for fixing vulnerabilities. In this work, we instead apply program repair techniques for smart contracts to fix exploits used by attackers. 

\vspace{5px}
\noindent\textbf{Fuzzing.} Fuzzing has been widely adopted in finding vulnerabilities\cite{aflpp,ossfuzz,clusterfuzz,fuzzfactory} and program repair\cite{fuzzingrepair}. Specifically, hybrid fuzzing\cite{hybrid0,hybrid1,hybrid2}, a combination of concolic execution and fuzzing, is leveraged to gain high test coverage over the program under test. Echidna\cite{echidna}, Harvey\cite{harvey}, ILF\cite{ilf}, sfuzz\cite{nguyen2020sfuzz} have been proposed for coverage-guided smart contract fuzzing. More recent works such as Smartian\cite{smartian} and ItyFuzz\cite{ityfuzz} identify the stateful nature of smart contracts and leverage state dataflow information to guide the fuzzing. 

\vspace{5px}
\noindent\textbf{Attack Detection.} Existing researches on detecting attacks leverage pattern matching~\cite{ad1,ad2,ad3,ad4eg} and large language model~\cite{ad5llm}. On the other hand, \sys is designed not to discern transactions and conduct analysis on every transaction. In a real-world deployment, to save computation resources, we use pattern matching, specifically \cite{ad1}, to filter benign transactions.

\section{Ethics Consideration}
\label{sec:ethical}
IRB has deemed this research not to be within the scope of human research.   
We have returned the funds and assets \sys rescued to the victims. This research does not involve finding or exploiting new vulnerabilities. 
Before sending the exploits generated by \sys to the chain, we try our best to reach out to the protocol developers. We would only send exploits under their permissions or when they do not respond after 2 hours. 
Every preemptive hijack attempt by \sys in the real world is used to counter-act a real attack from the hackers. 
All attack backrunning attempts are manually checked by one of the authors before \sys sending them to ensure they do not cause additional damage.   During the real-world experiment, \sys caused no collateral damage and rescued millions of dollars. 

The fuzzing and analysis process in \sys happens off-chain on our server and does not constitute a DoS attack on the network. 
All data used in this research are publicly available.  
After making this research work public, attackers may conduct adaptive evasion for \strategyone and \strategytwo. We mitigate this risk by avoiding sharing implementation details and configurations of \sys, making adaptive evasion hard. 

\section{Conclusion}
Our analysis demonstrates that existing attack frontrunning protections have become ineffective in real-world blockchain environments, with only 17 out of 158 attacks publicly visible for frontrunning. The prevalence of private transactions and intense competition between bots severely limit classic frontrunning approaches. To address these limitations, we propose two new automated defense strategies: \strategyone to protect potential victims before attacks by leveraging information from deployed exploits and \strategytwo to reuse attack transactions to safeguard similar contracts post-exploit. In a real-world deployment, \sys mitigated \totalrescuedamt attacks over two months to recover \totalrescued worth of assets, demonstrating the practical impact of our solutions.









    



    





    

\bibliographystyle{plain}
\bibliography{main}

\end{document}